\documentclass[12pt,preprint]{aastex63}
\usepackage{bbm}
\usepackage{mathrsfs}
\usepackage{amssymb,amsmath,float}
\usepackage{rotating}

\newcommand\ep{\epsilon}

\renewcommand\th{\theta}

\newcommand\ka{\kappa}
\newcommand\lam{\lambda}

\newcommand\om{\omega}

\newcommand\Om{\Omega}

\newcommand\<{\langle}
\renewcommand\>{\rangle}

\newcommand\ie{\emph{i.e.}}
\newcommand\eg{\emph{e.g.}}

\newcommand\beq{\begin{equation}}
\newcommand\eeq{\end{equation}}
\newcommand\bea{\begin{eqnarray}}
\newcommand\eea{\end{eqnarray}}
\newcommand\bal{\begin{align}}
\newcommand\eal{\end{align}}

\newcommand\fr{\frac}

\renewcommand\d{\mathrm{d}}

\newcommand\bj{\bold{j}}
\newcommand\bk{\bold{k}}

\newcommand\hbn{\hat{\bold{n}}}

\newcommand\br{\bold{r}}

\newcommand\bu{\bold{u}}

\newcommand\bA{\bold{A}}

\newcommand\bR{\bold{R}}

\renewcommand\bal{\mbox{\boldmath$\alpha$}}
\newcommand\bbe{\boldsymbol{\beta}}
\newcommand\dbbe{\dot{\boldsymbol{\beta}}}

\begin{document}

\title{Damping of long wavelength gravitational waves by the intergalactic medium}

\author{Richard Lieu}
\affil{Department of Physics and Astronomy, University of Alabama, Huntsville, AL 35899\\}
\author{Kristen Lackeos}
\affil{NASA Postdoctoral Program Fellow, NASA Marshall Space Flight Center, Huntsville, AL 35812\\}
\affil{Max-Planck-Institut f\text{\"u}r Radioastronomie (MPIfR), Auf dem H\text{\"u}gel 69, 53121, Bonn, Germany}
\author{Bing Zhang}
\affil{Department of Physics and Astronomy, University of Nevada, Las Vegas, NV 89154\\}

\begin{abstract}
The problem of radiation by the charged particles of the intergalactic medium (IGM) when a
passing gravitational wave (GW) accelerate them is investigated.   The largest acceleration
(taking a charge from rest to a maximum speed which remains non-relativistic in the rest frame of the unperturbed spacetime) is found to be
limited by the curvature of a propagating spherical gravitational wavefront. Interesting physics arises from
the ensuing emission of radiation into the warm hot IGM, which to lowest order is a fully ionized hydrogen plasma
with a frozen-in magnetic field $B$. It is found that for a vast majority of propagation directions, the right-handed polarized radiation can penetrate the plasma at
frequencies below the plasma frequency $\om_p$, provided $\om<\om_b,$ where $\om_b=eB/m_e$ satisfies
$\om_b<\om_p$ for typical IGM conditions. Moreover, the refractive index under such a scenario is $n\gg 1,$
resulting in an enhanced radiative dissipation of GW energy (relative to the vacuum scenario), which is more severe for electrons if both charge
species are in thermal equilibrium and accelerated in the same way. The emission by the electrons then prevails,
and is further amplified by coherent addition of amplitudes within the size one wavelength.
The conversion of GWs of $\lam\gtrsim 5\times 10^{13}$~cm to electromagnetic waves means such GWs
can only propagate a distance $\lesssim 1$~Gpc
before being significantly damped by an IGM B field of $\sim10^{-8}$ G.
The low-frequency GWs \textcolor{black}{targeted by pulsar-timing-arrays} will not survive unless the IGM magnetic field is much lower than expected.
The \textcolor{black}{mHz} frequency GW inspirals targeted by future \textcolor{black}{space based} detectors such as the Laser Interferometer Space Antenna
remain intact and can be detected.
\end{abstract}

\section{Introduction}

The interaction of GWs with matter and radiation of the Universe has been a subject of considerable interest for a long time.  A number of authors explored the observable consequences of the effect, in terms of \eg~the absorption of GWs by electron proton collisions \citep{fla19} or a collisonless but magnetized plasma \cite{ser01}, and scattering of GWs by a static magnetic field to produce radio waves \citep{mar00}.  Although it was found that GWs can propagate non-dispersively through a collisionless and unmagnetized plasma,
the \textcolor{black}{novelty} of this work is to show that so long as there are free charges with a frozen-in magnetic field, mode conversion from GW to electromagnetic radiation can take place, and the process prevents long wavelength GWs from propagating large distances, thereby limiting the visibility of some distant GW sources. \textcolor{black}{As noted by \cite{poi11}, a charged particle moving along the geodesic of curved spacetime does in general emit radiation. The radiation which results from the curvature of spacetime is indistinguishable from the physical acceleration of the charge with respect to the observer in Minkowski spacetime.}  

\textcolor{black}{While our work complements earlier papers on the question of absorption of GWs from the viewpoint of observability of distant GW sources, it also fills a significant gap in the literature, concerning the interaction of GWs with the IGM.} Attention has previously been focused on the role of viscosity, with \cite{haw66} showing that the attenuation is frequency independent. The follow-up work of \cite{zel83} concluded upon the insignificance of viscous effects at times later than the Planck time of the Universe.

\textcolor{black}{It should be noted that \cite{rev15} attempted a toy model for GW absorption with a charged gas cloud as the absorbing medium. In their work, gradients in the charge density lead to radiation. For our approach, plasma physics are used to model a realistic scenario (which crucially includes a frozen-in magnetic field) and draw observational consequences for cosmological GW propagation. The realistic astrophysical model we explore is GW propagation through the IGM. Below is the layout of the paper.}

\textcolor{black}{In Section 2, we explore how the peculiar velocity of a test mass modifies the GW acceleration. 
In that section, we also derive the upper bound on the area of a plane wavefront before wavefront curvature causes loss of transverse coherence. This limit is relevant to later sections. In Section 3.1, we show that quadrupole perturbations of a homogeneous unmagnetized plasma by a passing GW will not induce vector dipole currents of radiation emission when the ensemble of charges in question were initially at zero temperature. Electrons and positively charged ions will respond equally to a GW, and the total radiation will cancel identically, {\it unless} the difference in initial thermal velocities between the two species is taken into account. This asymmetry leads to a net emission of radiation. In Section 3.2 we introduce a frozen-in magnetic field and characterize the weakly anisotropic but strongly frequency dependent index of refraction for the medium. In Section 3.3 and 3.4, the radiation by an ensemble of charges is derived. In Section 3.5, we discuss coherent emission of long wavelength radiation by an ensemble of charges in a plasma, and find the radiative loss rate for electrons to be much higher than for protons. In Section 4 we integrate the {\it in situ} radiative dissipation of a propagating GW along our line-of-sight (of cosmological lengths) and explored the observational consequences.}

\textcolor{black}{Concerning the originality of our current work, there are three active ingredients that contribute to the paradigm of GW propagation over cosmological distance. The first is the recognition that the plasma of the IGM has a finite temperature, \ie~the charges have finite random velocities.  Secondly, we went beyond the analysis of \cite{rev15} by considering the realistic scenario of a {\it magnetized} IGM, which modifies the Maxwell tensor. Third is the integration of the emission rate, transversely across the area of negligible wavefront curvature (the maximum radius of Section 2) and longitudinally along the line-of-sight from source to observer.}

\textcolor{black}{We conclude the introduction with} a brief recapitulation of how a GW affects the relative position of particles floating in comoving space $(x,y,z)$ in the Transverse Traceless gauge, or TT-gauge.  For the $+$ polarization of GW propagation along the $z$ direction, the line element\footnote{This is the plane wave approximation of the spherical wave solution of the vacuum Einstein Field Equations. Validity of the approximation is given by (\ref{limitXY}).} is \beq ds^2 = c^2 dt^2 - (1+h) dx^2 - (1-h) dy^2 - dz^2 \label{ds} \eeq where $h=h(t,z)$ is the dimensionless wave strain.  The physical separation of a particle from the coordinate origin is given by $X=(1+h/2)x$,~$Y=(1-h/2)y$, and $Z=z$, for $|h| \ll 1$ and with $h$ being a function of time $t$ at any definite $Z=z$.  If
\beq h(t,z) = h_0 \fr{\lam_0}{z} \cos (k_0z-\om_0 t), \label{soln} \eeq where $h_0$ is a constant, $\om_0$ is \textcolor{black}{the angular frequency of the GW for non-evolving continuous waves from circular binaries (\eg~super massive black hole binaries that spiral inwards on timescales of thousands to millions of years, \cite{ses10b}; this is the type of sources whose observability are being investigated by this paper), and $k_0=\om_0/c$. For binary systems 
the GW frequency is twice the frequency of the orbit.} Equation (\ref{soln}) describes a plane GW (a good approximation far away from the source) propagating along the $z$-direction in the far field limit $z \gg \lambda_0 = 2\pi/k_0$.   The acceleration of the particle may be written as \beq \ddot X =  -\tfrac{1}{2} \om_0^2 h(t,z) X;~{\rm and}~\ddot Y = \tfrac{1}{2}\om_0^2 h(t,z) Y, \label{accel} \eeq provided one ignores terms involving $h^2$ (which includes the difference between $x$ and $X$, {\it etc.}, and peculiar velocity of the test mass, on the right sides of (\ref{accel})).  Since this work utilizes the thermal motion of charges as they are accelerated by a GW to calculate a net Poynting flux of radiation, (\ref{accel}) should strictly speaking include peculiar velocity terms.  In the next section, however, the effect of such terms are shown to be negligibly small indeed.

\section{Modification of GW acceleration by the peculiar velocity of the test mass}

Starting with the metric of (\ref{ds}), and (\ref{soln}), the physical distance between a test mass at comoving coordinates
$(x,y,z)$ and the origin is, to lowest order in $h$, given by  \beq X=\left(1+\fr{h}{2}\right) x;~Y=\left(1-\fr{h}{2}\right)y;~Z=z \label{Xx} \eeq when measured along the $x$, $y$, and $z$ directions.
If the test mass has no peculiar velocity, the first two time derivatives of $X$, $Y$, and $Z$ will be \beq \dot X= \fr{1}{2} \dot h x,~\dot Y = -\fr{1}{2} \dot h y, \dot Z=0;~\ddot X=\fr{1}{2} \ddot h x,~\ddot Y = -\fr{1}{2} \ddot h y, \ddot Z=0.
\label{dotdot} \eeq Thus (\ref{Xx}) and (\ref{dotdot}) imply, to $O(h)$, \beq \ddot X= -\fr{1}{2} \om_0^2 h X;~\ddot Y = \fr{1}{2} \om_0^2 h Y;~\ddot Z = 0, \label{ddotX} \eeq in agreement with
(3).

But will the agreement still hold when thermal peculiar motion of the test mass is taken into account?  When this motion is included, (\ref{dotdot}) is modified to become \beq \dot X= \fr{1}{2} \dot h x + \left(1+\fr{h}{2}\right)\dot x,~\dot Y = -\fr{1}{2} \dot h y + \left(1-\fr{h}{2}\right) \dot y,~\dot Z=\dot z. \label{dot} \eeq  Differentiating once more w.r.t. time, assuming no peculiar acceleration because thermal collisions are rare, one obtains \beq \ddot{X} = \fr{1}{2} \ddot h x + \dot h \dot x,~\ddot Y= -\fr{1}{2} \ddot h y- \dot h \dot y,~\ddot Z = 0. \label{ddot} \eeq  Now from (\ref{Xx}), one sees that if one replaces $x$ by $X$ {\it etc} in (\ref{dot}) and (\ref{ddot}), the error committed would only be $O(h^2)$.  Thus, to an accuracy $O(h)$ one may accordingly rewrite (\ref{dot}) and combine it with (\ref{soln}) to get  \beq \ddot{X} = -\fr{1}{2} \om_0^2 h X - \om_0 h \dot X,~\ddot Y= \fr{1}{2} \om_0^2 h Y + \om_0 h \dot Y,~\ddot Z = 0. \label{peculiar} \eeq  When compared to (\ref{ddotX}) and (\ref{accel}), one finds agreement between (\ref{peculiar}) and the other two equations apart from the $\dot X$ and $\dot Y$ terms. Thus, peculiar velocity does in principle alter the GW induced acceleration in the TT-gauge.

\textcolor{black}{In practice, however, let us examine the magnitude of the correction.
Since the GWs relevant to this paper have long wavelengths, we suppose
$\lambda_0 \approx 9 \times 10^{16}$~cm. This corresponds to an oscillation period $\approx$ 1 month, or $\om_0\approx 2.1 \times 10^{-6}$~rad~s$^{-1}$. At the end of this section we show, using (4), that gravitational plane wave approximation is lost beyond $X,Y \approx \sqrt{2\lam_0 z} \approx 2.4 \times 10^{22}$~cm for a source distance $z \approx 1$~Gpc. Considering this, we take $10^{22}$~cm as the typical value of $X$ and $Y$.
Next, we take the $\dot X$ peculiar velocity to be the r.m.s. speed of 0.1 keV electron in the IGM, which is $\approx 6 \times 10^8$~cm~s$^{-1}$.  Putting these numbers together, one finds
that the $\dot X$ and $\dot Y$ terms in (\ref{peculiar}) are
$\approx 4 \times 10^7$ times smaller than the metric terms. Namely it is acceptable to neglect the the $\dot X$ and $\dot Y$ terms in favour of $X$ and $Y$ terms.}  

\textcolor{black}{Note, the ratio of the two terms does not depend on $h$, but only scales with $\om_0$ and
source distance $z$ as $\sqrt{\om_0 z}$. One sees that even if $\om_0$ is as small as the Hubble
constant $H_0 \approx 2.3 \times 10^{-18}$~s$^{-1}$ (which means $z \approx c/H_0 \approx 1.3
\times 10^{28}$~cm), the contribution of any peculiar motion (at subluminal speeds as required by Special Relativity) to the GW induced acceleration remains negligible, relative to the metric contribution.}

\textcolor{black}{We also observe that, from (\ref{dot}) and (\ref{peculiar}), the time averaged rate of work done by a passing GW still vanishes. Namely, $\< \ddot X \dot X \> =0$, and the same for the other two directions, over an entire ensemble of test masses. This is the case even when peculiar motion is taken into account, provided that the ensemble average of the peculiar velocity vanishes by space isotropy.}

\begin{figure}
\begin{center}
\includegraphics[width=3in]{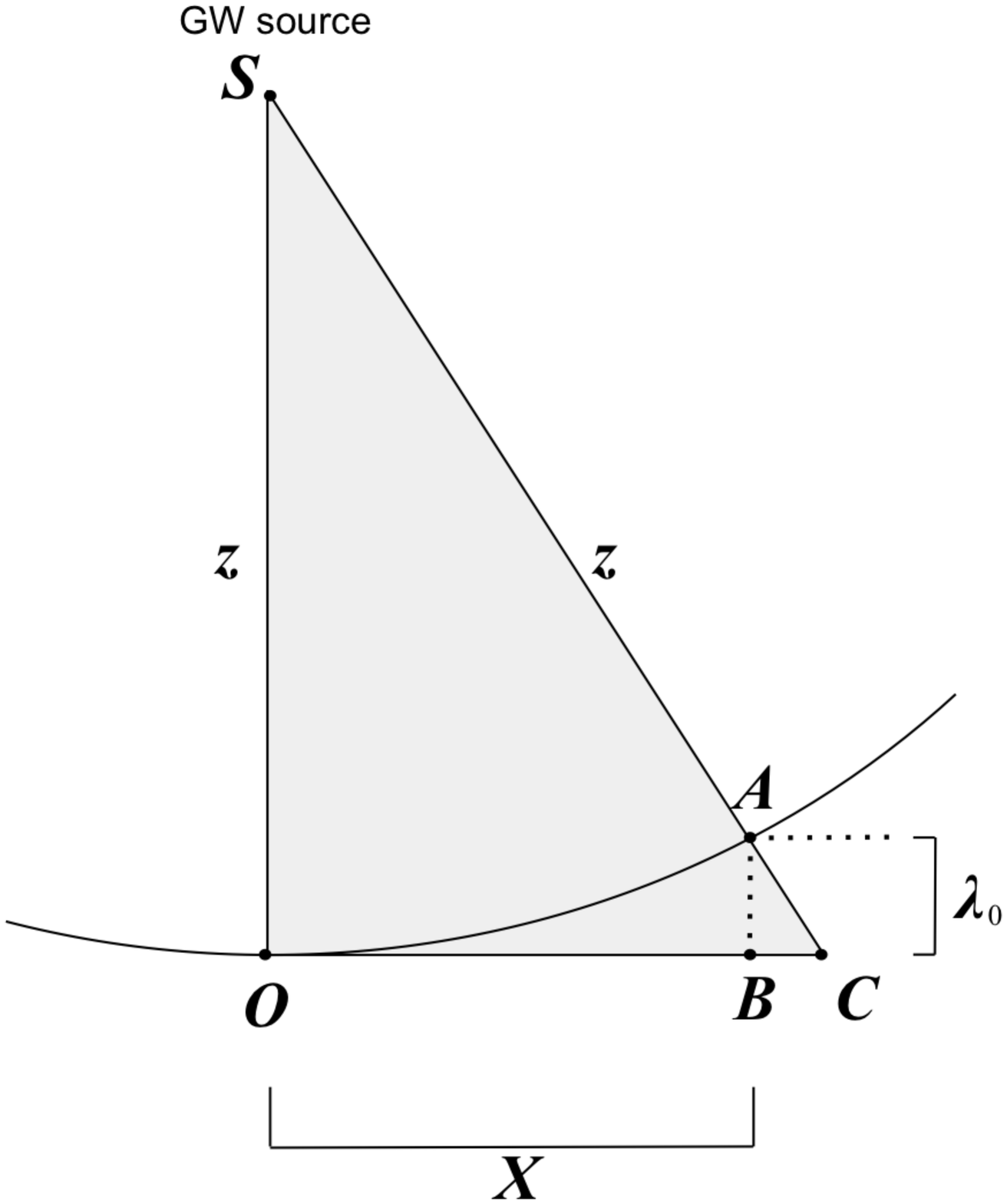}
\end{center}
\caption{The plane wave approximation of a spherical gravitational wavefront $OA$ is depicted $X$ as the solid line $OB$.
 If the separation $AB$ at the off-axis position $B$ is $\gtrsim \lam_0,$ the spatial coherence of $OA$ will not
 be perserved along $OB,$ and the approximation breaks down. For a distant point source $z\gg X\gg \lam_0$,
 $AC\simeq AB =\lam_0$, and the critical distance $X\simeq \sqrt{2\lam_0 z}$ marks the maximum spatial extent
 of the plane wave.}
\label{fig1}
\end{figure}

It is important to beware a criterion which stems from the realization that the line element (\ref{ds}) describes a plane GW propagating along $Z$, when GWs are usually emitted as spherical waves.  More precisely, the oscillation of an off-axis test mass is correctly described by (\ref{accel}) provided the phase of the plane gravitational wavefront at $(t,X,Y,Z)$ differs from $(t,0,0,Z)$, due to the finite radius of curvature $z$ of the wavefront, by an amount $\ll 2\pi$
(\ie\ provided the spherical wavefront reduces to a plane wave).  Simple trigonometry then converts the requirement to the inequality \beq |X|, |Y| \lesssim  \sqrt{2\lam_0 z}, \label{limitXY} \eeq where $\lam_0$ is the gravitational wavelength, see Figure \ref{fig1}.
We shall adopt (\ref{limitXY}) as the upper bound on $X$ and $Y$ for the coherent oscillation of a test mass under the influence of a GW. 
\textcolor{black}{This transverse length sets a limit to the plane wave approximation itself, and will be used in Section 4 to derive the total power converted from GWs to electromagnetic radiation, as they propagate through the IGM.}

\section{Radiation from a plasma accelerated by GW}

\subsection{Uncancelled gravitational-wave-induced electromagnetic radiation in vacuum}

One should begin by asking whether there is net electromagnetic radiation at all.
According to the Equivalence Principle the electrons and protons are accelerated by a GW in the same way, such that the wave fields from each species, being locally equal in magnitude, would cancel out each other everywhere.  This cancellation occurs, however, when an equal number of protons and electrons (assuming for simplicity a pure hydrogen plasma) within each volume of size $\lesssim\lambda^3$ are coherently accelerated and also move coherently in the same direction.\footnote{\textcolor{black}{The coherence volume is $\propto \lam^3$, because the emitting charges must be within one wavelength of the emitted mode for the ensuing radiation to be in phase in the far field.}}

\begin{figure}
\begin{center}
\includegraphics[width=4.5in]{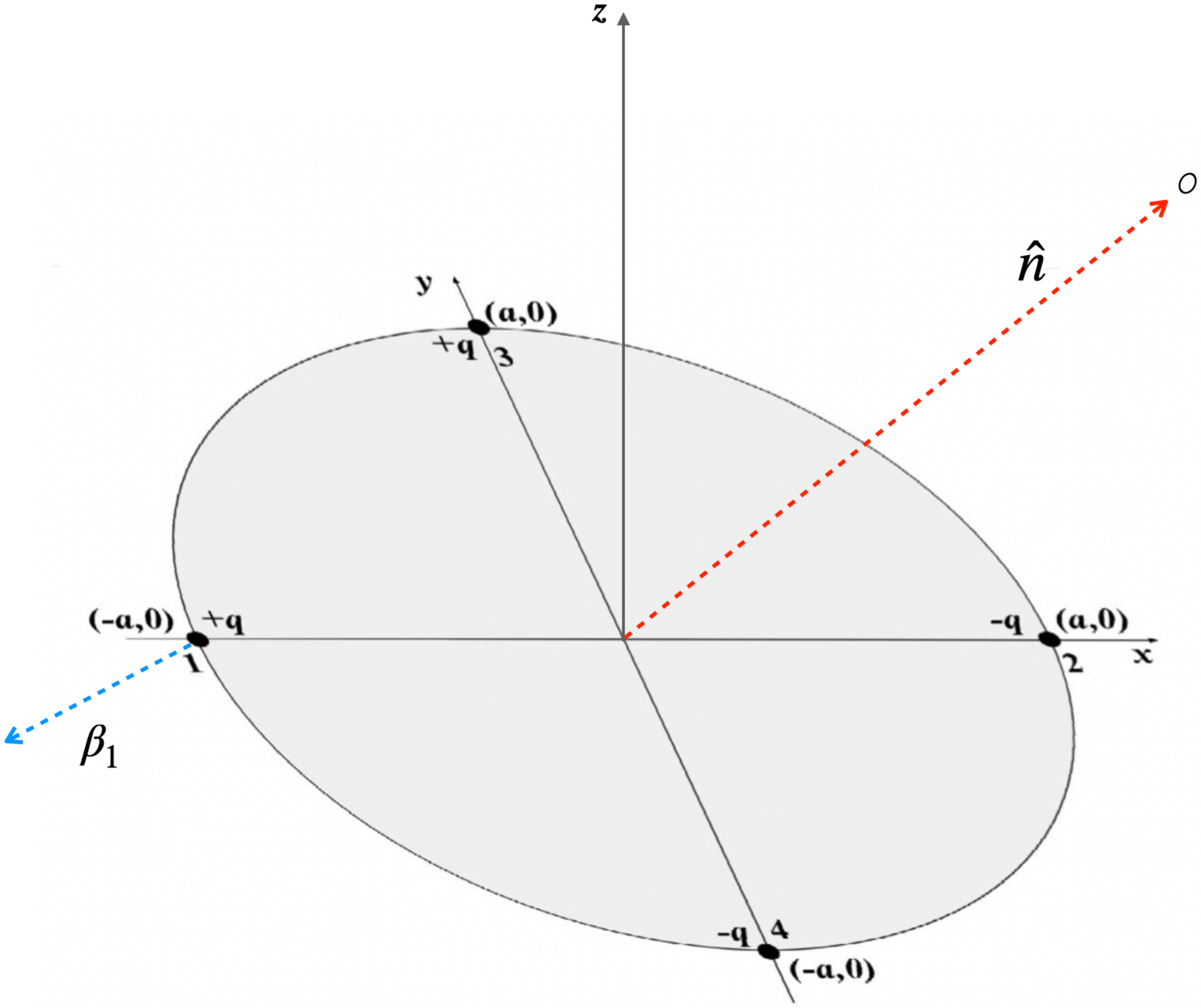}
\end{center}
\caption{Two pairs of charges $\pm$q are positioned at $(0,\pm a)\ {\rm and}\ (\pm a,0)$
 as a $+$mode GW propagating along the $z$-axis passes through.  The observer, $O$, is in the far field along some arbitrary direction
 in 3-space, and is given by the unit vector $\hat{\textbf{n}}$.  And $\bbe_j = {{\bf v}}_j$ depicts the thermal velocity of charge j in units of $c$.}
\label{fig3}
\end{figure}

\textcolor{black}{For a simple model of vacuum emission by an ensemble of charges, let us look at two pairs of identical test charges $\pm q$ positioned like the LIGO mirrors,
namely having separation $\sim 4$ km ($\ll$ gravitational wavelength). These are located at comoving coordinates $(0,a),\ (0,-a),\ (a,0),\ {\rm and}\ (-a,0)$ on the $xy$
plane as depicted in Figure \ref{fig3} (no need to enlist physical coordinates).} Let these charges be at rest in
Minkowski background spacetime. If there is no GW, there will obviously be no
radiation. \textcolor{black}{Consider a GW perturbation of the form}

\beq ds^2=c^2dt^2-(1+h)dx^2-(1-h)dy^2-dz^2,\ \ (\vert h\vert \ll 1).  \label{ap1}\eeq
The total vector potential of the radiation from the four charges in the far field and
dipole approximation limits (the latter requiring all four charges to be well within a space of size $\lam_0$, the former refers to emission to some distance $\gg\lam_0$) will be

\beq \bA=\fr{q}{4\pi\sqrt{\ep_0 c}}\sum^4_{j=1}\hat{\textbf{n}}\times\hat{\textbf{n}}\times\dot{\bbe_j}=0,\ \
\dbbe_{1,2}=\mp\fr{\om_0^2ha}{c}\hat{\textbf{i}}, \ \ \dbbe_{3,4}=\pm\fr{\om_0^2ha}{c}\hat{\textbf{j}}, \label{ap2}\eeq where $\hat{\textbf{n}}$ is the position vector of an observer in the far field.

\textcolor{black}{Now let each test charge have an initial velocity $\bbe_j = {{\bf v}}_j/c$, which is identical
among all four charges. The total vector potential becomes}
\beq \bA = \fr{e}{4\pi\sqrt{\ep_0 c}}  \sum^4_{j=1} \left[ \fr{\hbn\times\hbn\times\dbbe_j}{(1-\hbn\cdot\bbe)^{5/2}}
\right]=0. \label{ap3} \eeq In both cases, one does not require any delicate balance from the protons to
 prove that quadrupole perturbations cannot source any EM radiation.

Next, let the four charges possess initial velocities at {\it random} directions $\bbe_j$ (the meaning of $\bbe_j$ and $\hbn$ are further elucidated in Figure 2) which are uncorrelated with the GW acceleration directions (nor magnitudes), the
 total radiation amplitude becomes
 \beq \bA = \fr{e}{4\pi\sqrt{\ep_0 c}}  \sum^4_{j=1} \left[ \fr{\hbn\times\hbn\times\dbbe_j}{(1-\hbn\cdot\bbe_j)^{5/2}}\right]\neq 0. \label{ap4} \eeq
 In fact, it is possible to expand binomially the denominators to show that the
 vector potential becomes finite even to $O(v/c).$ Thus, in this simple heuristic
 picture, one can already see that random motion, including thermal motion, enables mode conversion of a GW
 to radiation within an ensemble of charges, even if the GW induces emission by
 accelerating all charges equally.

To be specific, the random motion particles in the ionized IGM have is thermal motion with velocities greatly exceeding the bulk motion velocity induced by gravitational waves. For the typical conditions of the warm intergalactic plasma\footnote{Only a pure hydrogen plasma is assumed. The number density of (\ref{IGM}) is obtained by assuming the baryonic IGM consists principally of a $10^5-10^7$ K plasma with normalized cosmic density $\Om_{\rm WHIM}\simeq$ 0.02 between z=0 and at least z=1 (Fig 2b \citep{cen99}), and $H_0=70$ km/s/Mpc.}, namely \beq kT \approx 0.1~{\rm keV};~n_e = n_i \approx 10^{-7}~{\rm cm}^{-3} \label{IGM} \eeq  the thermal velocity of electrons is $\<\beta_e^2\> = \<u_e^2\>/c^2 \approx 6 \times 10^{-4}$ and that of protons $\<\beta_p^2 \>$ is 1,836 times less, i.e. $\sim 3 \times 10^{-7}$. Both are much greater than the bulk motion velocity square $\beta_{\rm GW}^2 \sim (h \sqrt{2z/\lambda_0})^2 \sim 10^{-18}$ for typical values $z\sim 10^{28}$ cm, $\lambda_0 \sim 10^{16}$ cm, and $h \sim 10^{-15}$.
Under this circumstance, as was shown above, the wave field of
the electrons and protons will not cancel (because the former far exceed the
latter), and a net emission of radiation can be expected.

Thus, uncanceled residual excess emission arises from the $\beta_e$ correction term in the electron emission amplitude. (As emphasized above, peculiar velocities affect the overall acceleration negligibly when the metric is perturbed by a GW.) \textcolor{black}{In fact, for binary sources one can calculate the critical distance $z_c$ from the source, beneath which the GW dominates any thermal motion. Under such a scenario the GW induced mean square velocity exceeds $\<\beta_p^2\>$, and the difference in proton and electron thermal velocities might no longer be exploited to convert GWs to radiation.}  From equation (3.185) of \cite{cre11}, this distance is \beq z_c \approx 5 \times 10^{13} \beta^7 \left(\fr{10^6~{\rm K}}{T_{\rm IGM}}\right)
\left(\fr{\mu}{1~M_\odot} \right)~{\rm cm},\label{zc}\eeq where $\beta=v/c$ is the dimensionless orbital velocity and $\mu$ the reduced mass.  Such a  distance is an insignificantly small fraction of the propagation length.

\subsection{Propagation of gravitational wave induced radiation in the magnetized intergalactic medium}

\textcolor{black}{It is necessary to consider the emission of the accelerated charges to take place in a plasma rather than vacuum. The GW frequencies, hence the radiation they induce via charge acceleration in the IGM, are below the plasma frequency of the IGM. This means they cannot in general propagate through the ionized medium. However, in this section we show that when the role of a typical magnetic field which is frozen into the plasma, is taken into account, low frequency radiation emitted by the accelerated electrons will be able to propagate.}

\textcolor{black}{The all pervasive filamentary medium of the `IGM proper' space between clusters of galaxies, also known as the warm hot IGM, is critical to the paradigm of GW propagation over cosmological distances. From observations the frozen-in magnetic field has a strength of} \beq B\approx 10^{-8}~{\rm G}\label{B_IGM}\eeq in \cite{ak16}.
Thus, the \textcolor{black}{electron cyclotron frequency is
\beq \om_b=\fr{\left|e\right|B}{m_e}=0.176\left(\fr{B}{10^{-8}}~G\right){\rm rad\ s^{-1}} \label{cyclotron}\eeq}
On the other hand, for the intracluster filaments the plasma
temperature and density are given by (\ref{IGM}), and the
\textcolor{black}{plasma frequency given by \beq \om_p =\sqrt{\fr{n_ee^2}{\ep_0m_e}}= 17.8 \left(\fr{n_e}{10^{-7}~{\rm cm}^{-3}}\right)^{1/2}~{\rm rad~s}^{-1}. \label{omp2} \eeq}
Thus it is clear that \beq \om \ll \om_b \ll \om_p, \label{plasmacond}\eeq
for the GW sources we are interested in, where $\om$ is the GW frequency.
This inequality plays a major role in the propagation of GW induced radiation.

Unlike an unmagnetized plasma, very low frequency radiation satisfying the inequality (\ref{plasmacond}) and having a prescribed
polarization can propagate through a plasma with a frozen-in magnetic field.
Specifically, if one defines the three dimensionless parameters
\beq \xi=\fr{\om_p^2}{\om^2},\ \eta=\fr{\om_b^2}{\om^2}, \label{xi_eta}\eeq and if the angle
between the magnetic field and the radiation propagation direction is $\theta$,
the refractive index $n$ of the medium as given by \cite{jin06} simplifies, in the regime $\xi \gg \eta \gg 1$,
to the form
\beq n^2=1-\xi \left[1 +\fr{\eta\sin^2 \th}{2\xi} \pm \left( \fr{\eta^2\sin^4\theta}{4\xi^2}+\eta{\rm cos}^2 \theta \right)^{1/2}\right]^{-1} \label{etasq}\eeq

\textcolor{black}{Strictly speaking there is a contribution from the ions in the form of an additional term on the right side of (\ref{etasq}). For $\om_b \gtrsim\om \gg \om_{\rm bi} = ZeB/m_i$ this additional term is negligible. However, for $\om \ll \om_{\rm bi}$ it is identical in form to the $-\xi [\cdots]^{-1}$ part of (\ref{etasq}), except with the substitutions $\xi\to\xi_i = \om_{\rm pi}^2/\om^2 $, $\eta\to\eta_i = \om_{\rm bi}^2/\om^2 $, and $\pm\to\mp$.  For a pure hydrogen plasma the ion plasma and cyclotron frequencies, $\om_{\rm pi}$ and $\om_{\rm bi}$, are} readily evaluated by the replacement $m_e \to m_p$ in the original expressions for $\om_p$ and $\om_b$ (namely $\om_p \propto m_e^{-1/2}$ and $\om_b \propto m_e^{-1}$).  In the regime $\om\approx\om_{\rm bi}$, the expression for $n^2$ is more complicated than (\ref{etasq}) with the aforementioned extra term. This case is quite irrelevant because it applies to a very narrow spectral range, and because the emission of radiation at such low frequencies contributes negligibly to the total dissipation rate of a passing GW, as we shall see \textcolor{black}{in the remainder of this section}.

For plasma conditions satisfying (\ref{plasmacond}), we have \beq \xi \gg \eta \gg 1,\label{xi_eta_relation}\eeq
and (\ref{etasq}) reveals that apart from a narrow range of angles within the
interval \beq \delta \theta \lesssim \fr{\sqrt{\eta}}{2\xi} \leq 5 \times 10^{-5} \label{cone} \eeq on either side of
$\theta=\pi/2$, the cos$^2\theta$ term in (\ref{etasq}) dominates the sin$^4\theta$
and sin$^2\theta$ terms. That is, for the vast majority of propagation directions one
may ignore the sin$^4\theta$ and sin$^2\theta$ terms.  In this way, (\ref{etasq}) simplifies to \beq n^2=1-\fr{\om_p^2/\om^2}{1\pm\fr{\om_b}{\om}{\rm cos}\theta};~\om \ll \om_b \ll \om_p,~ \label{etasq_2}\eeq
where the $+$ and the $-$ sign correspond relatively to the left and the right-handed polarization
of the radiation respectively.  For ultra-low frequencies $\om \ll \om_{\rm bi}=\left|e\right|B/m_p$ then, in the case of a hydrogen plasma where 
$\om_{pi}=\sqrt{n_e e^2/(\ep_0 m_p)}$, 
\beq n^2=1-\fr{\om_p^2/\om^2}{1\pm\fr{\om_b}{\om}{\rm cos}\theta} -\fr{\om_{\rm pi}^2/\om^2}{1\mp\fr{\om_{\rm bi}}{\om}{\rm cos}\theta};~\om \ll \om_{\rm bi}~{\rm and}~\om_b \ll \om_p. \label{etasq_3}\eeq

For radiation frequency satisfying
(\ref{plasmacond}) with \textcolor{black}{the right handed polarization} mode propagating along directions $\cos\th >0$ and the \textcolor{black}{left handed polarization} mode along $\cos\th < 0$, (again, $\th$ well avoiding the very narrow `forbidden cone'  of $\left|\cos\th\right|\ll 1$ defined by (\ref{cone})), $n^2$ is $\gg 1$, and one may access the salient features of the propagation by setting $\left|\cos\th\right| \approx 1$ to simplify (\ref{etasq_2}) and (\ref{etasq_3}) to become, for $\om \lesssim \om_b$, \beq n \approx \fr{\om_p}{\sqrt{\om\om_b}} \th(\om - \om_{\rm bi}) + \fr{\om_p}{\sqrt{\om_{\rm bi} \om_b}} \th (\om_{\rm bi} - \om) \label{nrni} \eeq  where $\th (\om)$ is the Heaviside unit step function.  
\textcolor{black}{Note that the $\om\lesssim\om_{bi}$ limit is 
obtained by expanding (\ref{etasq_3}). Select the aforementioned choice of the $\pm$ 
sign for each mode and set $\left|\cos\th\right|\approx 1,$ and to first 
order of small $\om/\om_b$ and $\om/\om_{bi},$
$$n^2\simeq 1+\fr{\om_p^2}{\om\om_b}\left(1+\fr{\om}{\om_b}\right)-\fr{\om_{pi}^2}{\om\om_{bi}}\left(1-\fr{\om}{\om_{bi}}\right)$$
$$=1+\fr{\om_p^2}{\om_b^2}+\fr{\om_{pi}^2}{\om_{bi}^2}$$
$$\simeq \fr{\om_{pi}^2}{\om_{bi}^2}=\fr{\om_{p}^2}{\om_b\om_{bi}}.$$}
Thus the wavevector of the radiation is  \beq k = \fr{n_r\om}{c} \approx \fr{\om_p}{c} \left(\fr{\om}{\om_b}\right)^{1/2},~\om\ll\om_b\ll\om_p. \label{krki} \eeq    
Note also the invariance of the radiation propagation direction and polarization 
(both being defined with reference to the magnetic field orientation).

Evidently, based on the requirement given after (\ref{etasq_2}) the magnetoionic IGM is transparent to $\ka\approx 50$ \% of the unpolarized radiation induced by a passing GW emitted by an isotropic source at frequencies $\om \lesssim \om_b$.  With these realizations in mind, we shall henceforth assume that the $n$ is isotropic, and is given by the simplified version of (\ref{etasq_2}), \ie~(\ref{nrni}).
\textcolor{black}{In this section we have shown that the IGM with a frozen-in magnetic field is in principle transparent to a significant fraction of the unpolarized radiation induced by a passing GW. In the next two sections we explore the radiation process in practice.}

\subsection{Radiation by an ensemble of charges in the magnetized IGM plasma}
We first consider a single charge $q$ moving at velocity $\textbf{u}$, ($u\ll c$) as
it is being accelerated by a small amplitude wave of frequency $\om_0.$ \textcolor{black}{The position and velocity of the charge are given by \beq \br=\bu t+\bR\ {\rm sin}\om_0 t;\ \textbf{v}(t)=\textbf{u}+\om_0\textbf{R}{\rm cos}\om_0t. \label{rt_vt}\eeq
The background (thermal) velocity of the charge is $\textbf{u}.$ The passage of a GW of frequency $\om_0$ oscillates the charge at the same frequency. The amplitude of the oscillation of the charge is $|\textbf{R}|$, with the direction defined by the GW polarization.} The electric current density of the moving charge is
\beq \bj(\om,\bk)= q\int_{-\infty}^{\infty}dt\ \textbf{v}(t)\exp^{i(\om t-\textbf{k} \cdot \textbf{r})}.\label{current1}\eeq
To lowest order of $\textbf{k}\cdot \textbf{R},$ {\it viz.}~assuming
\beq kR \lesssim \fr{n(\om_b)\om_b}{c}h_0 \lambda_0^{3/2}z^{-1/2} \ll 1, \label{kR}\eeq
where $R=h_0\lambda^{3/2}z^{-1/2}$ from (\ref{soln}), (\ref{accel}) and
(\ref{Xx}), and $\om_b$ is the maximum frequency at which the
radiation can propagate because $n^2(\om)$ is positive (see (\ref{etasq_2}); note also we shall discuss in more detail after (\ref{c12}) why (\ref{kR}) is usually satisfied), the
current density is
$$ \bj(\om,\bk)= q\int_{-\infty}^{\infty}dt~\textbf{v}(t)\left(1-\textbf{k}\cdot
\textbf{R}\ {\rm sin}\om_0 t\right)\exp^{i(\om-\textbf{k} \cdot \textbf{v})t}$$
$$ = 2\pi q \Bigg\{\textbf{u}\delta(\textbf{k}\cdot \textbf{u}-\om)+\fr{1}{2}\left[\om_0\bf{
R}-\textbf{u}(\textbf{k}\cdot \textbf{R})\right]\delta(\textbf{k}\cdot \textbf{u}-\om-\om_0)+\fr{1}{2}\left[\om_0\bf{
R}+\textbf{u}(\textbf{k}\cdot \textbf{R})\right]\delta(\textbf{k}\cdot \textbf{u}-\om+\om_0)\Bigg\}.$$

The next step towards the radiation emission rate is to evaluate $\textbf{k}\times \textbf{j},$
as
\begin{multline}
\textbf{k}\times \textbf{j}=2\pi q\Bigg\{(\textbf{k}\times \textbf{u})\delta(\textbf{k}\cdot \textbf{u}-\om)+
\fr{1}{2}\left[\om_0(\textbf{k}\times \textbf{R})-(\textbf{k}\times \textbf{u})(\textbf{k}\cdot \textbf{R})\right]
\delta(\textbf{k}\cdot \textbf{u}-\om-\om_0)\\ +\fr{1}{2}\left[\om_0(\textbf{k}\times \textbf{R})+(\textbf{k}\times \textbf{u})(\textbf{k}\cdot \textbf{R})\right]
\delta(\textbf{k}\cdot \textbf{u}-\om+\om_0)\Bigg\}.
\label{kcrossj}
\end{multline}
The first $\delta$-function has to do with Cerenkov radiation (\cite{gri03}).
Moreover, the $\om_0 (\textbf{k}\times \textbf{R})$ term in the coefficient of the second
and third $\delta$-functions would dominate the $(\textbf{k}\times \textbf{u})(\textbf{k}\cdot \textbf{R})$
term if \beq \om_0 \gg \beta \om n(\om)\approx\beta\om_p\sqrt{\fr{\om}{\om_b}},\label{c5}\eeq
where the $\approx$ sign holds for $\om\lesssim \om_b.$ To begin with, we assume for
simplicity that (\ref{c5}) is satisfied, by ignoring the two $(\textbf{k}\times \textbf{u})(\textbf{k}\cdot
\textbf{R})$. This is the ultra low velocity limit.

The total emission energy (in Joules) is given by
\beq \mathcal{E}=\fr{1}{2\pi^3}\int_0^{\infty}\om d\om\int\fr{d^3\textbf{k}}{k^2}
{\rm Im}\left(\fr{\vert \textbf{k}\times \textbf{j} \vert^2}{k^2c^2-\ep\om^2}\right), \label{c6}\eeq
where Im(\dots) denotes the imaginary part, and
$\ep(\om,\textbf{k})=\ep_1(\om_1,\textbf{k})+i\ep_2(\om_2,\textbf{k})=n^2$ is the
dielectric constant of the medium into which the emission takes place.
Substituting (\ref{kcrossj}) into (\ref{c6}), one finds a spectral emission rate,
averaged over an isotropic distribution of charge velocities $\textbf{u},$ of
\begin{multline}  \fr{d^3\mathcal{E}}{dt d\om d\Om}=\fr{q^2\om}{4\pi^3}{\sin}^2\vartheta
\int^{2\pi}_0d\phi\int^{1}_0d{\rm cos}\theta \int^{\infty}_0\fr{k^2dk}{{\rm Im}(k^2c^2-\ep\om^2)}
\Bigg\{u^2\delta(\textbf{k}\cdot \textbf{u}-\om) \\+ \fr{1}{4}\om_0^2R^2\Bigg[\delta(\textbf{k}\cdot \textbf{u}-\om-\om_0)
+\delta(\textbf{k}\cdot \textbf{u}-\om+\om_0)\Bigg]\Bigg\}, \label{c7}\end{multline}
where the relation $\delta^2(\om)=T/(2\pi)$ with $T$ being the `total duration' of
emission was employed. It is also assumed that $(\theta,\phi)$ and $(\vartheta,\varphi)$
are respectively the polar and azimuthal angles that $\textbf{u}$ and $\textbf{R}$ make
w.r.t. a Cartesian system having $\textbf{k}\parallel \hat{\textbf{z}}$, and $d\Om=-d\varphi d{\rm cos}\vartheta.$

After performing the $k$ integration in (\ref{c7})
\begin{multline*} \fr{d^3\mathcal{E}}{dt d\om d\Om}=\fr{q^2\om}{4\pi^2 c^2}{\rm sin}^2\vartheta
  \int^1_0 \fr{d{\rm cos}^2\theta}{{\rm cos}^2\theta}\Bigg\{u{\rm Im}
  \left(\fr{1}{1-\ep\beta^2{\rm cos}^2\theta}\right)+\fr{1}{4}\fr{\om_0^2R^2}{u}
  \Bigg[{\rm Im\left(\fr{1}{1-\ep\beta^2d^2_+{\rm cos}^2\theta}\right)} \\
  + {\rm Im\left(\fr{1}{1-\ep\beta^2d^2_-{\rm cos}^2\theta}\right)}\Bigg]\Bigg\}
\end{multline*}

\begin{multline*}
  =\fr{q^2\om}{4\pi c^2}{\rm sin}^2\vartheta
  \int^1_0 d{\rm cos}^2\theta\Bigg\{\fr{u\Gamma}{\pi({\rm cos}^2\theta-{\rm cos}^2\theta_0)+\Gamma^2}
  +\fr{1}{4}\fr{\om_0^2R^2}{u} \Bigg[ \fr{\Gamma_+}{\pi({\rm cos}^2\theta-{\rm
  cos}^2\theta_+)+\Gamma^2}\\
  +\fr{\Gamma_-}{\pi({\rm cos}^2\theta-{\rm cos}^2\theta_-)+\Gamma^2} \Bigg] \Bigg\}
\end{multline*}

\begin{multline}
 =\fr{q^2\om}{4\pi c^2}{\rm sin}^2\vartheta \int^1_0 d{\rm cos^2\theta}\Bigg\{u\delta({\rm cos}^2\theta-
 {\rm cos}^2\theta_0)+\fr{1}{4}\fr{\om_0^2R^2}{u}\left[\delta({\rm cos}^2\theta-{\rm
  cos}^2\theta_+)+\delta({\rm cos}^2\theta-{\rm
  cos}^2\theta_-)\right]\Bigg\}\label{c8},
\end{multline}
where $d_{\pm}=\om/(\om\pm\om_0),$ $\Gamma=\ep_2/(\beta^2\vert\ep\vert^2)$,
$\Gamma_{\pm}=\ep_2/(\beta^2d_{\pm}^2\vert\ep\vert^2)$,
${\rm cos}^2\theta_0=\ep_1/(\beta^2\vert\ep\vert^2),$
${\rm cos}^2\theta_{\pm}=\ep_1/(\beta^2d_{\pm}^2\vert\ep\vert^2);$ and
the $\delta$-functions are the limiting case of no radiation absorption $\ep_2 \rightarrow 0$
 (hence $\Gamma \rightarrow 0$, which is the assumption we are
 making about the IGM).

\textcolor{black}{Before proceeding further with emission in a $\ep_2=0$, $\ep_1=n^2>1$ medium,
 let us perform a reality check by taking the $\ep_1=n^2=1$ limit. In this case,  $\delta({\rm cos}^2\theta-
 {\rm cos}^2\theta_0)=\delta({\rm cos}^2\theta-c^2/u^2)=0$, \ie~no Cerenkov radiation
 is possible. Also, $\delta({\rm cos}^2\theta-{\rm cos}^2\theta_+)=\delta({\rm cos}^2\theta-\beta^{-2}(1+\om_0/\om)^2)=0$, \ie~no negative frequencies $\om$ are possible, assuming $\om_0>0$.} 
\textcolor{black}{For $n=1$ the definition of $\cos\th_-$ as given immediately below (\ref{c8}), 
when used together with the 3$^{\rm rd}$ $\delta$-function of (\ref{c8}) after 
it is integrated over $\cos^2\th$, yields $\om_0=\om(1-ux/c)$ where $x=\cos\th_-.$ 
Thus $\om{\rm d}\om=u\om_o^2/[c(1-ux/c)^3],$} which leaves us with
 \beq \fr{d^3\mathcal{E}}{dt d\Om}=\fr{q^2\om_0^2R^2}{16\pi c^2u}{\rm sin^2}\vartheta \int^{\om_{\rm max}}_0\om d\om=
 \fr{q^2\om_0^4 R^2{\rm sin}^2\vartheta}{16\pi c^3}\int^1_{-1}\fr{dx}{(1-\fr{u}{c}x)^3}. \label{c9}\eeq
 \textcolor{black}{In the limit $u\rightarrow 0,$ $d\mathcal{E}/dt=q^2\om_0^4R^2/(3c^3).$ This is
 consistent with the non-relativistic Larmor formula} $d\mathcal{E}/dt=2q^2\< \dot{\textbf{v}}^2\>/(3c^3)$
 with $\textbf{v}(t)$ as given by (\ref{rt_vt}) ({\it viz.}~$\< \dot{\textbf{v}}^2\>=\om_0^4R^2\<{\rm sin}^2\om_0t\>=
 \om_0^4R^2/2.$)

\textcolor{black}{For the near isotropic magnetized plasma of the IGM, which is transparent to the 
right-handed polarized radiation at low frequencies, (\ref{nrni}) yields the $\ep_2=0,$ $\ep_1=n^2\gg 1$ scenario 
for 50$\%$ of unpolarized radiation.
This is applicable to practically all directions of propagation.} Once again, we shall ignore the first $\delta$-function of (\ref{c7}).

In the limit of $n\beta \ll 1$, an inequality consistent with (\ref{c5}), the second and third
$\delta$-functions of (\ref{c7}) ensure that $\om\simeq \pm \om_0.$ Only the third
$\delta$-function which enforces $\om\simeq \om_0$ is acceptable, and
(\ref{c7}) becomes
\beq \fr{d\mathcal{E}}{dt}=\fr{q^2\om_0^4R^2n(\om_0)}{3c^3},
\label{c11}\eeq
which is the Larmor formula enhanced by the factor $n$.

\textcolor{black}{In the next section we turn to the question of the radiative loss rate of the electrons relative to the protons of the IGM when (\ref{c5}) is violated. In this regime, an asymmetry exists between the two particle species, that sets the stage for GW absorption to take place.}

\subsection{GW induced emission of the IGM}
The inequality of (\ref{c5}) is violated by both electrons and protons (apart from those irrelevantly\footnote{\textcolor{black}{Irrelevant because when $\om$ is small enough to satisfy (\ref{c5}) and thereby reinstate the dominance of the $\om_0k\times R$ term of (\ref{kcrossj}), coherent emission of radiation by IGM electrons under the influence of a passing GW cannot take place, see the beginning of Section 3.5.}} small frequencies, $\om\ll\om_0$), as one can easily check by
consulting (\ref{IGM}), (\ref{cyclotron}), (\ref{omp2}) and
(\ref{etasq_2}) for the conditions of the IGM. This means the $(\textbf{k}\times \textbf{u})(\textbf{k}\cdot \textbf{R})$
term in the coefficient of the last two $\delta$-functions of (\ref{kcrossj})
dominates the $\om_0 (\textbf{k}\times \textbf{R})$ term for the problem of interest.
This term now replaces the $\om_0\textbf{R}$ term for the case studied in Section 3.3.
Moreover, the inequality (\ref{kR}) which ensures the sufficiency of expanding the exponential of (\ref{current1}) to only the $\textbf{k}\cdot \textbf{R}$ order is
satisfied provided
\beq h_0 \ll 99.4 \left(\fr{\lam_0}{5.2\times 10^{13}{\rm cm}}\right)^{-3/2}
\left(\fr{z}{3\times 10^{27}{\rm cm}}\right)^{1/2}. \label{c12}\eeq
 The limit required by (\ref{c12}) is well above the typical $h_0$ expected for GWs of wavelength 
 $\lam_0\simeq$ $ 5\times 10^{13}$ cm, see {\it e.g.}~(\cite{jar09}). 
 
\textcolor{black}{Within the two coefficients of (\ref{kcrossj}) one ignores the $\om_0\textbf{R}$ term in favor of $(\textbf{k}\times \textbf{u})(\textbf{k}\cdot \textbf{R})$, and the calculation proceeds along the same lines as Section 3.3, to arrive at a total loss rate of}
 \beq \fr{d\mathcal{E}}{dt}=\fr{e^2}{2c^2u}R^2\beta^2\int^{\om_b}_0 n^2(\om)\om^3 d\om=
 \fr{e^2R^2\om_p^2\om^2_b u}{6c^4}, \label{c13}\eeq where use was made of (\ref{nrni}) in the last step.
The loss rate is considerably less for protons than electrons because protons have much smaller equipartition velocities at any given IGM temperature.

 \textcolor{black}{Specifically, the conditions of the IGM as given by (\ref{IGM}), (\ref{cyclotron}), (\ref{omp2}) and
(\ref{etasq_2}) yielded a r.m.s. thermal velocity of $\beta_e\gtrsim 10^{-2}$ $\beta_p\lesssim
 10^{-3}$. From (\ref{etasq_3}) a refractive index of $n\simeq 10^2$ at
 the upper integration limit of $\om =\om_b$ is given, where $\om_b$ is
 given by (\ref{cyclotron}). Beyond this limit $n(\om)$ is negative.} For the second and third $\delta$-functions of (\ref{c8})
 one may write, with the help of (\ref{etasq_3}),
 \beq {\rm cos}\theta=\fr{1}{n\beta}\left(1-\fr{\om_0}{\om}\right)=\fr{\sqrt{\om_b}}{\om_p\beta}
 \fr{\om-\om_0}{\sqrt{\om}}.\label{c14}\eeq 
 \textcolor{black}{We assume that $\om_b\gg\om_0$, bearing in mind this paper is
 primarily about low frequency GWs, $\om_0\lesssim 10^{-6}{\rm rad~s}^{-1}$. Then a maximum of $\om_{\rm max}=\om_b$
 is evidently accessible from electron emissions. While for protons, (\ref{c14})
 indicates ${\rm cos}~\theta\simeq 10$ already at $\om=0.1\om_b$. This
 clearly indicates $\om_{\rm max}\ll\om_b$.} In fact, with $\om_b$
 and $\om_p$ given by (\ref{cyclotron}) and (\ref{omp2}) respectively, $\om_{\rm max}\simeq 10^{-3}\om_b$
 for protons. \textcolor{black}{Thus, revisiting (\ref{c13}), the $\beta\om_{\rm max}^2$ scaling of $d\mathcal{E}/dt$ indicates that the radiative loss rate is much higher for electrons than protons.}

 Turning to the problem of emission by an ensemble of charges, the first
 important point to note is again the comparison between electron and proton emission.
 For vacuum emission $n=1$, where (\ref{c9}) reduces to the standard Larmor
 formula for $d\mathcal{E}/dt$, it is clear that to lowest order in $\beta=u/c$
 electrons and protons have the same radiation power when they are subject to
 the same acceleration, as in the case of a passing GW. Since the two charge
 species have $q=\pm e,$ a small volume of fully ionized pure hydrogen plasma
 within which the emission amplitudes are in phase will undergo destructive
 interference resulting in no emission at all on average.  \textcolor{black}{The purpose of this section was to demonstrate, in contrast, the asymmetry in thermal velocities introduces a higher order effect in $\beta$, which one can see in (\ref{c13}), explicitly.}

\subsection{Coherent emission}

Specifically one can write the net current as
\beq \textbf{j}(\om,\textbf{k})=\sum^N_{l=1}e\int \textbf{v}_{\rm pl}(t)\exp^{i(\om
t-\textbf{k}\cdot\textbf{r}_{\rm pl})}dt
-\sum^N_{m=1}e\int \textbf{v}_{\rm em}(t)\exp^{i(\om t-\textbf{k}\cdot\textbf{r}_{\rm em})}dt, \label{c15}\eeq
where $\textbf{r}_{\rm pl}=\textbf{u}_{\rm pl}t+\textbf{R}{\rm sin}(\om_0 t)+\textbf{R}_{\rm pl};\
\textbf{r}_{\rm em}=\textbf{u}_{\rm em}t+\textbf{R}{\rm sin}(\om_0 t)+\textbf{R}_{\rm em},$
with the suffixes $p$ and $e$ denoting protons and electrons respectively. The
difference between (\ref{c15}) and (\ref{current1}) is the existence of the phases $\textbf{k}\cdot\textbf{R}_{\rm pl}$
and $\textbf{k}\cdot\textbf{R}_{\rm em}$ in (\ref{c15}). 
 \textcolor{black}{The terms in each of the two summations that contribute to the constructive interference at a far field position produce a finite contribution to $\textbf{j}(\om,t)$. The contributing terms have phases that differ from each other by an amount $\ll 2\pi$ within each set.}
 \textcolor{black}{Since $k=n\om/c,$ this requires the position of the charges to occupy a coherence volume $\lesssim \lam^3 = (2\pi/k)^3=8\pi^3c^3/(n^3\om^3)$.  Moreover, there is also the additional requirement of $\lam\ll$ the GW wavelength $\lam_0=2\pi c/\om_0$ (or equivalently $n\om\gg\om_0$). Otherwise the emitting electrons within the $\lam^3$ volume would not be accelerated in tandem under the influence of the passing GW. For GWs with 
$\om_0\lesssim$ 1 mHz (frequencies central to this paper), it is easily verified that the 
$\lam\ll\lam_0=2\pi c/\om_0$ inequality ensures the violation of (\ref{c5}), hence the validity of keeping the $(\textbf{k}\times \textbf{u})(\textbf{k}\cdot \textbf{R})$ term of (\ref{kcrossj}) in preference to the $\om_0 (\textbf{k}\times \textbf{R})$ term (see the beginning of Section 3.4).}

Ignoring the absolute phase $\textbf{k}\cdot\textbf{R}_0$ which disappears upon taking
$\vert \textbf{k}\times \textbf{j}\vert^2$, and assuming (\ref{current1}), (\ref{c15})
becomes, for a maximum volume of coherent emission

\begin{multline} \bj(\om,\bk)= 2\pi e\sum^{N(\om)}_{l=1} \Bigg\{\textbf{u}_{\rm pl}\delta(\textbf{k}\cdot \textbf{u}_{\rm pl}-\om)+\fr{1}{2}\om_0\bf{
R}\left[\delta(\textbf{k}\cdot \textbf{u}_{\rm pl}-\om-\om_0)+\delta(\textbf{k}\cdot \textbf{u}_{\rm
pl}-\om+\om_0)\right]\Bigg\}\\ -
2\pi e\sum^{N(\om)}_{m=1} \Bigg\{\textbf{u}_{\rm em}\delta(\textbf{k}\cdot \textbf{u}_{\rm em}-\om)+\fr{1}{2}\om_0\bf{
R}\left[\delta(\textbf{k}\cdot \textbf{u}_{\rm em}-\om-\om_0)+\delta(\textbf{k}\cdot \textbf{u}_{\rm
em}-\om+\om_0)\right]\Bigg\},\label{c16}
 \end{multline}
where \beq N(\om)=\fr{8\pi^3c^3n_e}{n^3\om^3}.\label{c17}\eeq
 \textcolor{black}{In the limit $n=1,$ and $\beta_{\rm pl}=n u_{\rm pl}/c \ll 1$ for all $l$ (protons) 
 and likewise for all $m$ (electrons), the emission frequency is $\om=\om_0$.
Specifically, one can ignore $\textbf{k}\cdot\textbf{u}_{\rm pl}$ and $\textbf{k}\cdot\textbf{u}_{\rm em}$
in the arguments of the 3$^{\rm rd}$ and 6$^{\rm th}$ $\delta$-functions, in which case the two summations
in (\ref{c16}) cancel each other. This results in no emission under
this non-relativistic vacuum scenario. Even more precisely, this cancellation applies to lowest order in 
$\beta=u/c$,~\ie~for vacuum emission $n=1$ and to
order of $\beta$ the radiative power is at $\om=\om_0$ and stems from the third
and last $\delta$-function of (\ref{c16}) with $\textbf{u}_{\rm pl}=\textbf{u}_{\rm em}=0$
for all $l$ and $m$.}  \textcolor{black}{The second and fifth $\delta$-functions yield negative frequencies (and can be ignored).
The 1$^{\rm st}$ and 4$^{\rm th}$ are Cenenkov terms which this work does not address because the effect
 has nothing to do with any passing GWs.}

The above analysis may further be elucidated and extended by contemplating the
more general possibility of $n>1$ but $n\beta\rightarrow 0$ such that (\ref{c5})
is satisfied. Consider the scenario relevant to the current problem of
propagation of low frequency radiation in a magnetized plasma. If $n\beta\ll 1$
and $\beta\ll 1$ for all the charges, the arguments which led to (\ref{c11})
have revealed that the lowest order contribution to $\textbf{j}(\om,\textbf{k})$
would still be coming from the third $\delta$-function in each of the
last $m$ summations of (\ref{c16}) with $\textbf{u}_{\rm pl}=\textbf{u}_{\rm em}=0$
for all $l$ and $m$. This once again means $\textbf{j}(\om,\textbf{k})$ vanishes
to lowest order of $\beta,$ leading to no emission to this order.

Yet the situation changes when $n$ and $\beta$ are slightly larger, such that (\ref{c5})
is violated despite $\beta\ll 1$. Under this scenario, which is more relevant to
the problem of GWs propagating in the IGM, (\ref{c13}) reveals that electrons
radiate far more efficiently than protons, and one may ignore the latter.

Thus, taking into account only the electrons and bearing in mind the violation
of (\ref{c5}) means the $(\textbf{k}\times\textbf{u})(\textbf{k}\cdot\textbf{R})$
term of (\ref{kcrossj}) now dominates over the $\om_0\textbf{k}\times\textbf{R}$ term, (\ref{c16})
becomes for an ensemble of many electrons (and ignoring the Cerenkov term)
\begin{multline}
\left\vert\sum_p \textbf{k}\times\textbf{j}_p\right\vert^2=
8\pi^2e^2\sum^{N(\om)}_{p\neq q}\fr{1}{4}
(\textbf{k}\cdot\textbf{R})^2(\textbf{k}\times\textbf{u}_p)\cdot(\textbf{k}\times\textbf{u}_q))
\Bigg[\delta(\textbf{k}\cdot\textbf{u}_p-\om-\om_0)\delta(\textbf{k}\cdot\textbf{u}_q-\om-\om_0)
\\
+\delta(\textbf{k}\cdot\textbf{u}_p-\om+\om_0)\delta(\textbf{k}\cdot\textbf{u}_q-\om+\om_0)\Bigg]
\label{c18}
\end{multline}
Substituting into (\ref{c6}), then the integral over $k$ becomes
\begin{multline}\int dk \fr{\delta(\textbf{k}\cdot\textbf{u}_p-\om-\om_0)\delta(\textbf{k}\cdot\textbf{u}_q-\om-\om_0)}
{{\rm Im}(1-\ep\om^2/c^2k^2)}=\fr{1}{u_p{\rm cos}\theta_p}{\rm Im}
\left(\fr{1}{1-\ep\beta_p^2d_+^2{\rm cos}^2\theta_p}\right)\\
\times\delta\left(\fr{u_q{\rm cos}\theta_q}{u_p{\rm
cos}\theta_p}(\om+\om_0)-\om-\om_0\right)\label{c19}\end{multline}
for the first $\delta$-function of (\ref{c18}), with $\om_0\rightarrow-\om_0$ for
the second $\delta$-function, where $d_{\pm}$ are defined after (\ref{c8}). Now the
$\delta$-function of (\ref{c19}) is survived (\ie~it becomes $T/2\pi$) by a
fraction $\approx1/\pi$ of the $N(\om)$ electrons in the $q$-summation, for
which $\theta_q=\theta_p$ but $\phi_q$ is independent of $\phi_p$ (we assumed for simplicity that $u_q=u_p$
for all $q\neq p$). There is a second solution of $\theta_q=\pi-\theta_p$ which
yields much less radiation power, at the single frequency $\om=\om_0,$ which
will not be discussed.

In this way, after averaging over an isotropic distribution of $\textbf{u}_l$
and $\textbf{u}_m$, one obtains
\begin{multline} \fr{d^3\mathcal{E}}{dt d\om d\Om}=\fr{e^2\om}{8\pi^2c^2}N^2({\om}){\cos}^2\vartheta
\int^1_0(d{\rm cos}^2\theta){\rm \tan}^2\theta R^2 \Bigg[(\om+\om_0)^2\delta({\rm cos}^2\theta-{\rm
cos}^2\theta_+)\\
+(\om-\om_0)^2\delta({\rm cos}^2\theta-{\rm cos}^2\theta_-)\Bigg]\label{c20}\end{multline}

Upon integration w.r.t. ${\rm cos}^2\theta$ one obtains
\beq\fr{d^3\mathcal{E}}{dt d\om d\Om}=\fr{e^2R^2\om}{8\pi^2c^2u}N^2({\om}){\cos}^2\vartheta
\left[2n^2\beta^2\om^2-(\om+\om_0)^2-(\om-\om_0)^2\right].\label{c21}\eeq
Since $n\beta\gg 1$ for electrons, and $n\om\gg\om_0$, where $\om\gtrsim\om_0$, because (as discussed above)
most of the radiative power is at frequencies $\om\simeq\om_b\gg\om_0$, one may
ignore $\om_0$ in (\ref{c21}) and write, after summing all $\textbf{k}$
directions,
\beq\fr{d^2\mathcal{E}}{dt d\om}=\fr{e^2R^2u\om^3}{3\pi c^4}N^2(\om)n^2(\om).\label{c22}\eeq
Lastly, we integrate over $\om$ by using (\ref{nrni}) and (\ref{c17}) to get
\beq\fr{d\mathcal{E}}{dt}\Big\vert_V=\int^{\om_b}_0\d\om\fr{d^2\mathcal{E}}{dtd\om}\fr{V}{\lam^3}=
\fr{16\pi^2e^2uR^2\om^2_bn_e^2V\kappa}{9\om_pc},\label{c23}\eeq
which is the total radiative power over a large volume $V\gg\lam_b^3,$ where $\lam_b=2\pi c/(n\om_b)$

(wavelengths
much longer than $\lam_b$ are irrelevant as they contribute negligibly to the
integral). Note the quantity $\kappa=0.5$ in (\ref{c23}) serves as a reminder
that only the right-handed polarization component of the GW induced radiation
can propagate through the magnetized IGM, see the previous subsection.

\begin{figure}
\begin{center} 
\includegraphics[width=7in]{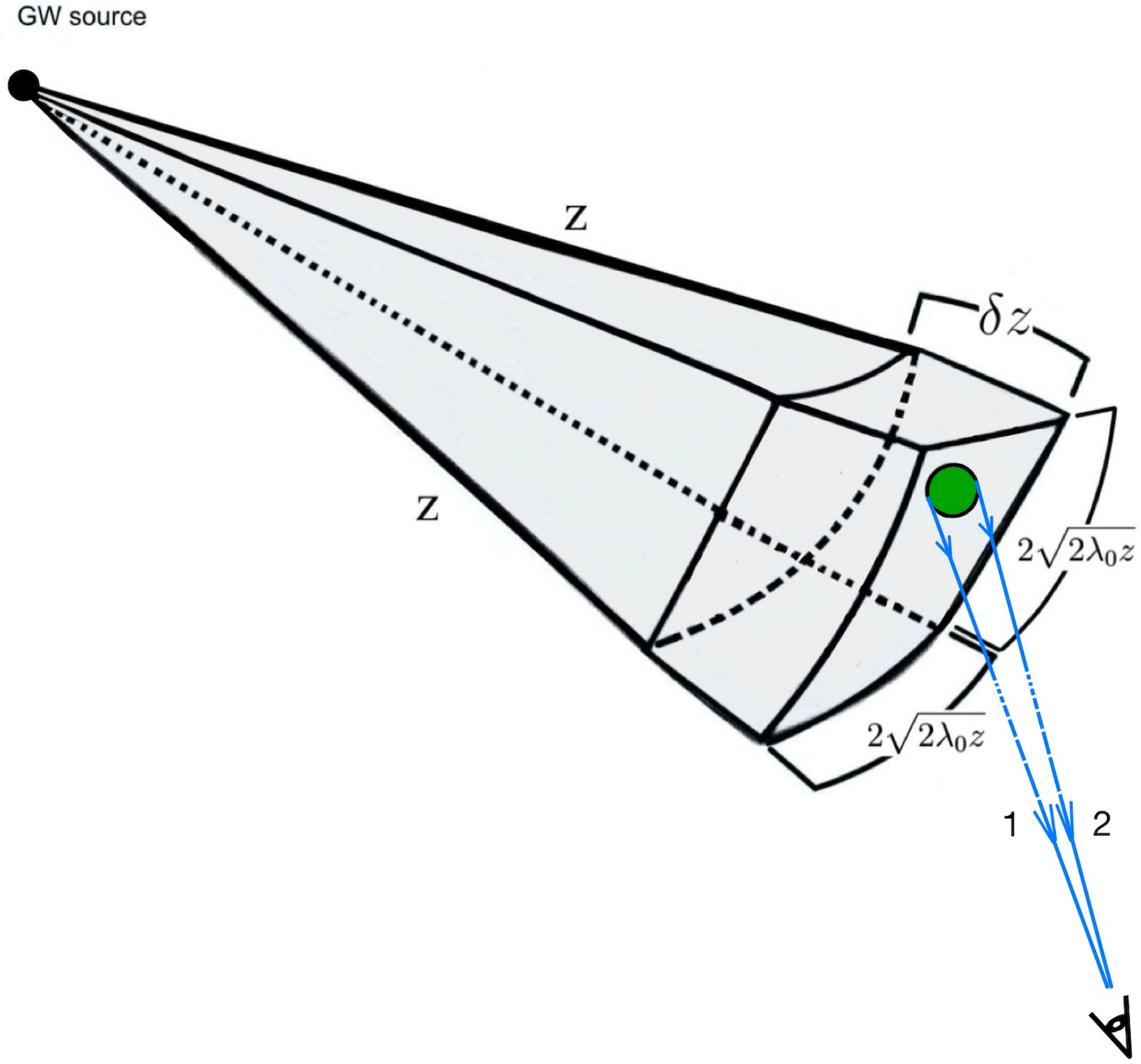}
\end{center}
\caption{A slab of GW induced radiation and thickness $dz,$ containing many coherence volumes of radius
$\approx \lam$ (small circle inside slab) is shown as part of an entire spherical shell of radius $z$.
The slab area is defined by the phase invariance of the plane wave approximation of the GW, and sets 
the radius
of the slab at $\sqrt{2\lam_0z}\gg \lam_0.$  \textcolor{black}{Within each volume $\approx\lam^3$} the emission vector 
amplitudes
are in phase at the far field (see subsection 3.3, especially immediately after (\ref{c15})) 
so that the total brightness of the volume is proportional to the {\it square} of the number of electrons in it.
\textcolor{black}{The amplitudes of waves emitted by charges located within the green colored sphere of size $\lam$ 
are in phase at the far field point, having propagated along paths 1 and 2, which differ in pathlength by $\ll \lam$.}
From one such volume to another, intensities
add to form the total loss rate of each slab, \ie~the total intensity of all the volumes is proportional to the 
number of volumes and not the square of it.}
\label{slab}
\end{figure}

\textcolor{black}{At this point, we remind the reader of two important length scales in this paper. The plane wave approximation scale (\ref{limitXY}) (also see Figure \ref{fig1}) is distinct from a secondary scale which defines the boundary to a group of charged particles emitting coherent electromagnetic radiation. This secondary scale $\sim\lam$ is much smaller than one gravitational wavelength, $\lam_0$. Namely, charges within a region whose maximum size is no larger than $\lam \ll \lam_0$ will emit EM waves in phase at the far field. Such regions are spatially contiguous and extend transversely (to the the GW propagation direction) to the plane wave limit (\ref{limitXY}). Figure \ref{slab} serves as a guide to these two length scales.}

\section{The damping length of cosmological GW emissions}

With the results of the previous section, it is now possible to estimate the damping length of GWs in the intergalactic medium.   Consider a slab of plasma along the line-of-sight to a steady GW source.  Let the distance of the slab from the source be $z$, and the thickness of the slab be $dz$.   For the $+$ mode of GW propagation, the acceleration at physical distance $X$ from the slab axis is given by the first equation of (\ref{accel}).  More generally, for any polarization mode, the acceleration at cylindrical radius $\rho$ has square magnitude $\approx \om_0^4 h^2 \rho^2/4$ where $h=h_0\lam_0\cos (\om_0 t - kz)/z$, and mean square magnitude $\om_0^4 h_0^2 \lam_0^2 \rho^2/(8z^2)$, extending by (\ref{limitXY}) to $\rho_m = \sqrt{2\lam_0 z}$.  The total number of such slabs which form a spherical shell of radius $z$ and centering at the source is \beq N_{\rm slab} \approx \fr{4\pi z^2}{\pi\rho_m^2}. \label{Nslab} \eeq    As a check, note that the product of the volume of each slab, $\rho d\rho d\phi dz$,  and the number of slabs comprising the shell gives, upon integrating from $\rho=0$ to $\rho=\rho_m$, the total volume $4\pi z^2 dz$ of the shell.

One may proceed to calculate the radiation power of one plasma slab as it is accelerated by the GW.  By (\ref{c23}), it is
\beq d^3P_{\rm slab} = \fr{d\mathcal{E}}{dt}\Bigg\vert_{\delta V} \delta V = \fr{8\pi^3 e^2}{9} \left(\fr{3kT}{m_e}\right)^{1/2} \fr{n_e^2\om_b^2 h_0^2 \lam_0^2 \rho^3\kappa}{\om_p c z^2} \delta \rho \delta z.  \label{Pslab} \eeq  Multiplication by $N_{\rm slab}$ of (\ref{Nslab}), followed by integration from $\rho=0$ to $\rho=\rho_m$, where $\rho_m$ is given by (\ref{limitXY}), then gives the power emitted by the entire spherical shell of radius $z$ and thickness $dz$ from below. After that, integration from $z=0$ to $z=L$ where $L$ is the physical separation between the source and the observer (in this treatment one ignores the expansion of the Universe), yields the total power converted from GW to radiation by the time the GW has propagated through a column $L$ of intergalactic medium.  The result is a rate of conversion from GW to electromagnetic power given by
\beq P_{\rm EM}  = \fr{8\pi^3}{9} \left(\fr{3kT}{m_e}\right)^{1/2} \fr{e^2 n_e^2 \om_b^2 h_0^2\lam_0^3 L^2\kappa}{\om_p c}. \label{Pem} \eeq  For comparison, we express the total GW luminosity of the source in terms of the same quantities.  Starting with the GW energy density\footnote{Strictly speaking there is no gauge invariant expression for the GW energy density, and the expression we use to derive (\ref{Pgw}) is valid and meaningful only if $\lam_0$ the gravitational wavelength is much smaller than the curvature of the background spacetime.  In the present context of cosmological propagation, this criterion is satisfied provided $\lam_0 \ll c/H_0$ where $H_0$ is the Hubble constant.} at any point along the line-of-sight, $U_{\rm GW} = c^2 \om^2 |h|^2/(32\pi G)$ \citep{har03} where (consistent with (\ref{soln})) $|h|=h_0 \lam_0/z$, and noting that the flux is $U_{\rm GW} c$ and hence the total emission rate is $4\pi L^2 U_{\rm GW} c$, one finds
\beq P_{\rm GW} = \fr{\pi^2}{2} \fr{c^5}{G} h_0^2. \label{Pgw} \eeq
The ratio of the radiation loss rate to the GW emission rate \beq \fr{P_{\rm EM}}{P_{\rm GW}} = 1.00 \left(\fr{kT}{0.1~{\rm keV}}\right)^{1/2} \left(\fr{n_e}{10^{-7}~{\rm cm}^{-3}}\right)^{3/2} \left(\fr{L}{1~{\rm Gpc}}\right)^2 \left(\fr{\lam_0}{5.2 \times 10^{13}~{\rm cm}}\right)^3\left(\fr{B}{10^{-8} G}\right)^2\left(\fr{\kappa}{0.5}\right) \label{quotient} \eeq does not depend on the intrinsic strength $h_0^2$ of the GW source.
From (\ref{quotient}), it can be seen that over a propagation length of $L \approx 1$~Gpc, all the GW emission is converted to electromagnetic energy via the acceleration  of electrons in the ionized warm intergalactic medium for gravitational wavelengths longward of $5\times 10^{13}$~cm, or frequencies below $\om_0=$ 3.8~mHz.

Could the absorption be compensated by stimulated emission along the same
direction \cite[e.g.][]{fla19}? While \cite{fla19} discussed the
stimulated emission of GWs being compensated by absorption of GWs, the current
work is about the absorption of electromagnetic radiation after they are mode converted from an
incident GW beam. Thus, if there is any stimulated emission, it will be in the
form of electromagnetic radiation and not GWs, \ie~one does not envisage  such a compensation of GWs to take place.

\section{Conclusion and discussion}

Gravitational waves with wavelength $\lam_0 \gtrsim 5\times 10^{13}$ cm are shown to be significantly attenuated by the IGM as they propagate distances $\gtrsim$ 1 Gpc. The GW is converted into electromagnetic waves which heat the surrounding plasma. A difference in the r.m.s. thermal velocities of the protons and electrons within the IGM leads to coherent plasma emission and a corresponding reduction in the GW energy.

\begin{figure}[htbp]
\centering
\includegraphics[width=6.0in]{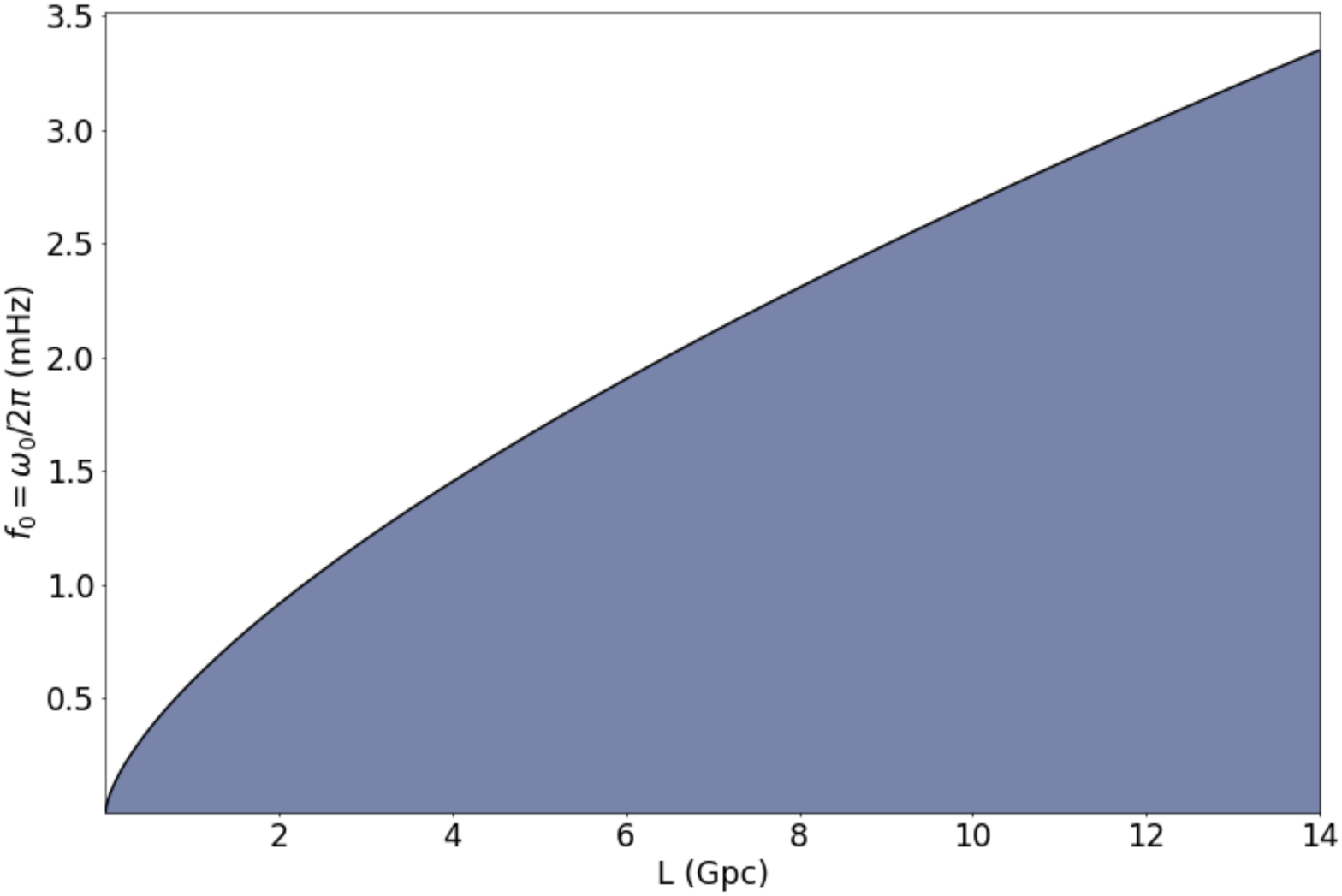}
\caption{The black line represents the frequency where $P_{\rm GW}=P_{\rm EM}$ as a function
of distance from the source $L$ and source frequency $\om_0/2\pi$. The $x$-axis spans the range 1 kpc to 14 Gpc. The solution uses $B=10^{-8}G$
 and $n_e=10^{-7} {\rm cm}^{-3}.$
GWs with frequencies below the cutoff line (shaded region) are absorbed by the IGM.}
\label{figtwo}
\end{figure}

In Figure \ref{figtwo}, we plot the locus of $P_{\rm GW}=P_{\rm EM}$ on the $(\om_0, L)$ plane, where $\om_0$ is the GW frequency and $L$ is the GW
attenuation length as given by (\ref{quotient}).
Continuous GWs from early inspiral massive black hole binaries (MBHB,
with mass $\sim 10^9 M_{\odot}$ (\cite{min17}, \cite{az14})
fall below the cutoff line.
 GWs from these sources, anticipated to have frequencies in the range of $1-1000$ nHz and
 are targets for pulsar-timing-array GW detectors \citep{spo19}, are entirely absorbed by the IGM with a frozen-in magnetic field.
 Gravitational waves from MBHB systems with masses $10^4-10^7 M_{\odot}$ at the stage of coalescence and merger
 will have higher frequencies which survive IGM absorption. Stellar and intermediate mass BBHs (\cite{ses17}, \cite{jan20})
 will also largely survive absorption. These events will still be observable by space GW missions such as the Laser Interferometer Space Antenna (LISA) (\cite{am17}) \textcolor{black}{in the mid to high frequency range of the full LISA band spanning 0.1 mHz to 1 Hz.}

One aim of pulsar timing observations is to detect stochastic and continuous GWs sources in the 1-1000 nHz
frequency range. The Pulsar Timing Arrays (PTA) search for the stochastic GW background has 
produced upper limits on a population of supermassive black hole binaries (SBHB) \citep{agg19,len15,sha15,arz16, arz18, arz20b}. 
PTA derived limits for individual SBHB sources have also been produced \citep{ses09,ses10,sch15,bab16,arz20a}.  
At even lower frequencies new detection methods \citep{bat10, yon16} have been used to constrain, 
`ultra-low-frequency GWs' from the Galactic Center \citep{ku19} and M87 \citep{ki21}.

We note that GW-diminishent by GW absorption has been mentioned in a 
PTA-related context before. For example, in the pulsar timing analysis work of \cite{sha15} 
the authors suggest their limit is consistent 
with GW absorption on cosmological scales and reference \cite{haw66} as an explanation. \cite{spo15}
also mention GW absorption as an effect that may diminsh the GW signal, and 
commented on the implausibility of the viscosity driven mechanism treated by \cite{haw66}.

There is recent news from PTA groups on their search for a stochastic GW background. 
The NANOGrav collaboration found Bayesian evidence for a `common-spectrum stochastic process' 
in its 12.5-yr PTA dataset; and more recently the EPTA collaboration also report the detection of a 
common red noise signal, when analyzing a timespan up to 24 years \cite{ch21}. 
However, so far, no group has found significant evidence of the quadrupolar Hellings $\&$ Downs 
inter-pulsar spatial correlations \citep{arz20b}, and no individually resolvable 
GW sources in the PTA band have been detected.
Moreover, \cite{gon21} showed it is possible to spuriously detect a common red 
process in timing array data set simulation when no such process is injected into the simulated data.

There is growing observational evidence that SBHB at sub-parsec and parsec separations exist, where candidates are identified through periodic light curve variability
\citep{sud03, era12,gr15,ch16, liu16, bas17,bri18,ser20,kov20}. One candidate SBHB is
excluded by PTA measurments \cite{zhu18}. \textcolor{black}{Another study by \cite{ha18} looked at quasi-periodic BL Lacertae objects and flat-spectrum radio quasars from the {\it Fermi} Gamma-Ray Space Telescope. They showed that if these quasi-periodic candidates were all binaries, the stochastic background from such a putative population would be well above the limits placed by PTA's, see their figure 3. From there, the authors constrained the fraction of blazars hosting a binary with orbital periods $<$5 yr, to less than $10^{-3}$, to maintain consistency with PTA limits. Additional assumptions and modeling are typically required to evolve SBHB candidates into the PTA frequency band to determine if new candidates are in agreement with PTA limits \cite{guo19}.} Though no additional candidates are excluded by current PTA limits, SBHB gravitational wave background estimates at nHz frequencies (\cite{ses18}) are beginning to be in tension with pulsar timing upper limits.

The result presented in this paper offers an explanation to pulsar timing `non-detections' of 
individually resolvable, circular SBHBs. 
In turn, a minimum magnetic field strength is inferable for the distances within which SBHBs have been ruled out by pulsar timining upper limits. Take \cite{agg19} as one example. 
The authors exclude SBHBs with $\mathcal{M} > 1.6\times 10^9{\rm M}_{\odot}$ 
emitting GWs with $\omega_0$ = 17.6-1996.7 nHz in the Virgo Cluster. 
Assuming a distance of 20 Mpc, we use equation 6 of \cite{ade16} to estimate an electron number density 
of $\sim 2.3\times 10^{-6}~{\rm cm}^{-3}$. Using Eq (\ref{quotient}), we find the mean magnetic field strength required to 
absorb GWs of frequency 1997 nHz emitted from the center of the Virgo Cluster is no more than $\simeq 6\times 10^{-13}$ 
G, and for 17.6 nHz is $\approx$ 1200 times smaller. The magnetic field strength 
of the Virgo Cluster is estimated to be on the order $\mu$G \citep{pf10}, so 
it remains plausible the magnetic field of the Virgo Cluster would significantly 
disrupt nHz GW propagation emitted from any source inside. 

\begin{figure}[h!]
\begin{center}
\includegraphics[width=6in]{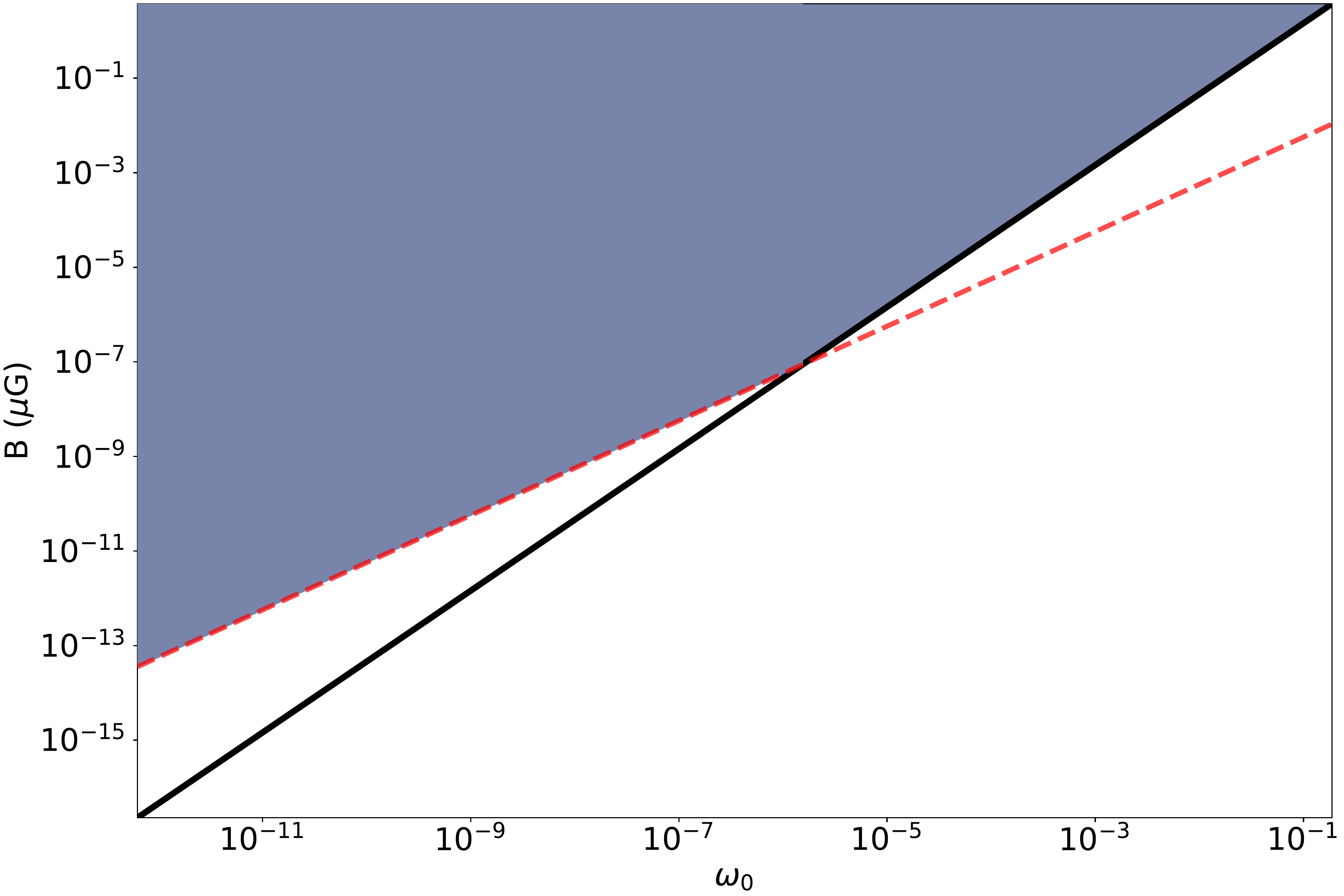}
\end{center}
\caption{The solid line is the locus of $P_{\rm GW}=P_{\rm EM}$ on the $(B, \om_0)$ plane, for 
GW sources emitting at a distance 1 Gpc, with an electron number density of $n_e=10^{-7} {\rm cm}^{-3}$.
The dashed line is the locus for condition $\om_0<\om_b$, $B>e\om_0/m_e$. 
GWs traveling in a plasma with a frozen-in magnetic field 
that falls outside of the shaded area will survive absorption.}
\label{minB}
\end{figure}

For GW sources at the distance of 1 Gpc, the magnetic field as a function of GW frequency 
satisfying $P_{\rm EM}=P_{\rm GW}$ is shown as a solid line in Figure \ref{minB}. For abosorption, the 
GW is subject to the additional constraint that its frequency $\om_0$ is less than the cyclotron frequency $\om_b$. 
This constraint is indicated by the dashed line. The region where both conditions are satisfied is shaded.  
Outside this region, GWs will travel through the magnetized IGM unabsorbed.

\textcolor{black}{We end with a summary of the key assumptions and approximations used. First, a charged particle accelerated gravitationally can emit radiation at a rate $\sim \dot u^2$, Section 1. Second, the GW is emitted by a distant and steady source, such that its amplitude $h\sim\cos(k_0z-\om_0 t)/z$, where $z$ is the distance to the center of emission, (\ref{peculiar}). Third, the peculiar velocity of test masses was shown after (\ref{peculiar}) to play a negligible role in determining the acceleration by a passing GW, even though it is so much greater than the r.m.s. velocity due to the wave. Fourth, the plane wave nature of a gravitational wavefront breaks down beyond a radius of the polarization field $\sim \sqrt{2\lam_0z}$. Fifth, the GW angular frequency $\om_0$ is assumed to be well beneath the cyclotron frequency $\om_b$ of electrons in the frozen-in magnetic field of the IGM, which in turn is well beneath the electron plasma frequency $\om_p$ of the IGM. Sixth, 50$\%$ of the radiation emitted by such electrons at frequencies $\om<\om_b$ (when they are accelerated by the GW) can propagate through the IGM, experiencing a large, real, and positive refractive index $n$. This 50$\%$ comprises right-handed polarization for emission along the magnetic field and left-handed polarization for emission anti-parallel to the field. Once this is taken into account, space isotropy in other respects are mild and can be ignored. Seventh, the standard formula for the spectral angular distribution of radiation emitted by an accelerated charge in a medium of $n\gg1$ indicates, in the context of acceleration by a charge by a passing GW, that the $(\textbf{k}\times \textbf{u})(\textbf{k}\cdot \textbf{R})$ contribution to the emission amplitude far exceeds, the lowest order (dipole) contribution $\om_0 (\textbf{k}\times \textbf{R})$. Eighth, the peculiar velocity difference between thermal electrons and protons of the IGM prevent complete cancellation of the total amplitudes from the two charge ensembles. Ninth, the emission amplitudes of electrons within a volume $V$ of size $\approx \lam,$ where $\lam$ is the radiation wavelength, reinforce each other in the far field to result in a $N^2\sim n_e^2\lam^6$ dependence of the emission rate, provided $\lam<\lam_0$ (which is not equivalent to $\om>\om_0$, because $n\gg1$), so that the charges are accelerated in lockstep by the GW.}

\textcolor{black}{Finally, this paper is about the absorption of GWs from a distant point source by the 
IGM. It does not address the fate of a stochastic GW background of a population of MBHBs (\cite{ses08}, \cite{rav15}), 
which could result from the absorption and subsequent remission (\ie~scattering) of GWs from point 
sources by the IGM, thereby removing their direction information but not necessarily the GW 
power they emitted. Such a topic is outside the scope of this paper.}

\newpage

KL's research was supported by an appointment to the NASA Postdoctoral Program at the NASA Marshall Space Flight Center,
administered by Universities Space Research Association under contract with NASA.

\end{document}